# Holographic dark information energy: predicted dark energy measurement.

Michael Paul Gough
Department of Engineering and Design, University of Sussex,
Brighton, BN1 9QT, UK
E-mail: m.p.gough@sussex.ac.uk

**Abstract.**
Several models have been proposed to explain the dark energy that is causing universe expansion to accelerate. Here the acceleration predicted by the Holographic Dark Information Energy (HDIE) model is compared to the acceleration that would be produced by a cosmological constant. While identical to a cosmological constant at low redshifts, $z < 1$, the HDIE model results in smaller Hubble parameter values at higher redshifts, $z > 1$, reaching a maximum difference of 2.6 ± 0.5% around $z$ ~1.7. The next generation of dark energy measurements, both those scheduled to be made in space (ESA's Euclid and NASA's WFIRST missions) and those to be made on the ground (BigBOSS, LSST and Dark Energy Survey), should be capable of determining whether such a difference signature exists or not. The HDIE model is therefore falsifiable.

**1 Introduction**
The expansion of the universe is accelerating, driven by a dark energy that presently accounts for around three quarters of the total energy of the universe. This conclusion, initially obtained using type 1a supernova as reference sources [1][2], is supported by more recent supernova measurements [3] and now confirmed by a number of independent measurements [4-7]. The large dark energy component is evident in gravitational weak lensing [4], in the projected scale of baryon acoustic oscillations [5], in the Cosmic Microwave Background radiation anisotropies [6], as well as in the growth rate of large scale structure, or clustering power spectrum of galaxies [7] (see reviews [8][9] ). The many explanations proposed to explain dark energy include: Einstein's cosmological constant; some form of quintessence field; and a wide range of 'alternative' models [8][9].

Two properties are required for a dark energy model to fit the observations. Firstly, the model must be capable of quantitatively accounting for the dark energy density value, a high value, three times the energy density equivalence of the universe's total mass. Secondly, at least for the recent period, redshifts $z<1$, the model must provide a near constant energy density, equivalent to a total dark energy that increases as ~$a^3$ where $a$ is the universe scale size (size relative to today, $a=1$). This corresponds to a dark energy equation of state value $w$~−1, since energy densities vary as $a^{-3(1+w)}$. Ideally we wish to find a model that satisfies these two requirements without recourse to exotic or unproven physics.

Foremost amongst likely explanations is the cosmological constant, or vacuum energy, that satisfies the second of our two requirements, by definition exhibiting a constant energy density, equivalent to the specific equation of state value $w=-1$. In contrast, quintessence is a scalar field with a dynamic equation of state that varies over space and time. Experimental measurements [10] at low redshifts, $z < 1$, limit the dark energy equation of state to lie within the narrow range $w=-0.94\pm0.09$ and thus generally favour the cosmological constant explanation. Accurate measurements of dark energy at higher redshifts, $z>1$, await the next generation of measurements. Unfortunately, satisfying the first requirement is more difficult as there are only two possible energy density values expected for a cosmological constant from quantum field theories [11]. The cosmological constant should either have a preferred value some 120 orders of magnitude higher than that required - a value impossible to reconcile with our universe, or it has the value zero. Then a different dark energy explanation would, if verified, enable the cosmological constant to take this second natural value: zero.

A number of holographic dark energy models attempt to account for dark energy without invoking exotic particles, exotic fields, modifications to gravity, or interactions with dark matter [12-14]. The anti-de-Sitter/conformal field theory duality leads directly to considering that all of the information describing a system can be considered as being encoded on the system's bounding surface [15][16]. Then today's dark energy density can follow from a combination of the Planck scale with a suitable cut-off dimension, for example the infrared cut-off [17]. Holographic dark energy may also be caused by vacuum entanglement [18], effectively energy from quantum information loss via Landauer's principle [19].

Holographic dark energy models [12-14],[17][18] generally aim to provide an all pervading vacuum energy in the form of a cosmological constant. In contrast the specific model considered in this paper is the Holographic Dark Information Energy, HDIE, model [20] which takes a phenomenological approach. Here dark energy is explained as the energy equivalence of the information, or entropy, associated with stars and stellar heated gas and dust. HDIE accounts for both of the required dark energy properties and, in particular, manages to satisfy the first property using well established physics (see discussion of section 2.2). Nevertheless, HDIE is only one of the many dark energy models proposed to date and science progress requires that we eliminate all but one. With that elimination in mind, the emphasis here is on the HDIE model's predicted signature that differentiates HDIE from other models/theories.

## 2 The HDIE model
The HDIE model has been described in detail before [20] and only relevant features are recounted here. Essentially, HDIE combines Landauer's principle [19] with the Holographic principle [15][16].

Landauer's principle [19][21-23] states that any 'erasure' of information, or reduction of information bearing degrees of freedom, requires a minimum of $k_B T \ln 2$ of heat per erased bit to be dissipated into the surrounding environment. This dissipated heat increases the thermodynamic entropy of the surrounding environment to compensate for the loss of degrees of freedom and comply with the 2nd law. Information is not destroyed as the 'erased' information is now effectively contained in the extra degrees of freedom created in the surrounding environment.

Heat dissipation from information erasure is comparatively weak and usually insignificant in our normal day to day experience. For example, world-wide, man-kind has now accumulated some $10^{22}$ bits of stored digital data and we have the technological capacity to process a total of around $3 \times 10^{19}$ instructions per second in general-purpose computers [24]. We can assume that the main information erasure that occurs during the process of computing is caused by the overwriting the processor's instruction register when each new instruction is read from memory. In this way man-kind erases some $10^{21}$ bits each second. At room temperature this rate of information erasure will generate a world-wide total of only 3W! This is insignificant, ~$10^{-11}$ of the total electronic heat dissipation (ohmic and inductive heating, etc) of the world's ~$10^9$ computer systems that each dissipate ~$10^2$W. Similarly, erasing the $10^{22}$ bit sum total of all man made stored digital data would generate a world-wide total of just 3J.

Despite this low bit equivalent energy, the information-to-energy conversion process has now been demonstrated experimentally using Brownian particles under feedback control [25] and by a one-bit memory consisting of a single colloidal particle trapped in a modulated double-well potential [26]. Moreover, Landauer's heat dissipation from information erasure is still considered the best way to reconcile Maxwell's Demon with the second law of thermodynamics (see reviews [23] and [27] and references therein).

Landauer's principle provides the information energy equivalence, similar to the $mc^2$ energy equivalence for mass. When the same degrees of freedom are considered information entropy and thermodynamic entropy are identical. Then every component of the universe has an information equivalent energy of $Nk_BT\ ln\ 2$ that depends on the quantity of information (or entropy), $N$ bits, associated with the component and on the component's temperature, $T$. In this way we can consider the energy represented by information in the cosmos without requiring, or even identifying, processes whereby information may be actually 'erased'.

The Holographic principle [15][16] states that the amount of information in any region, $N$, scales with that region's bounding area. The Holographic Principle lead directly from the discovery that the maximum entropy of a black hole is set by its surface area [28] but the principle is considered to have universal validity [29], i.e. not just limited to the maximum entropy limit of black holes.

**2.1 Stellar heated gas and dust**

Stellar heated gas and dust and black holes have been found to make the greatest contributions to the entropy, $N$, of the universe [30][31][32]. Furthermore, stellar heated gas was estimated [20] to have the highest $NT$ product, making the largest information energy contribution to the universe (see Table 1 of [20]). Black holes could make the next strongest contribution at a few percent of that level but it is doubtful whether the information within a black hole, and therefore its information energy, has any effect on the universe because of the 'no hair theorem' [33]. While a black hole may exert a significant gravitational force on local objects, the only information that the universe has about it is limited to just three parameters: mass; charge; and angular momentum. From the universe's information point of view a black hole is no more than just another single fundamental particle, albeit a massive one!

Since the information energy, $Nk_BT\ ln2$, of the universe is primarily determined by stellar heated gas and dust, the appropriate temperature, $T$, will be the average temperature of baryons in the universe. Figure 1(a) plots average baryon temperature, $T$, data and the fraction of baryons in stars, $f$, deduced from a wide literature survey of integrated stellar density measurements, extending the earlier HDIE work [20]. Data symbols and measurement source references are listed here: open circle [34]; open squares [35]; filled rectangles [36]; diamonds [37]; upside down triangles [38]; normal triangles [39]; crosses [40]; circles with dot [41]; filled circles [42] and blue line [43].

Figure 1(a) shows that the average temperature of baryons today is $T\sim2\times10^6$ K, which, together with the estimate of $N\sim10^{86}$ from surveys [30][31][20], provides a quantitative estimate of the present HDIE energy value within an order of magnitude of the observed dark energy, satisfying dark energy requirement 1. Note that this is dependant on our estimate of $N$ for stellar heated gas and dust, only accurate to a couple of orders of magnitude. We find that, despite the very low bit equivalent energy, information energy can provide a significant contribution on cosmic scales, primarily because the universe's mass has remained constant while both the quantity of information (entropy), $N$, increased continually and the average baryon temperature, $T$, also increased with increasing star formation.

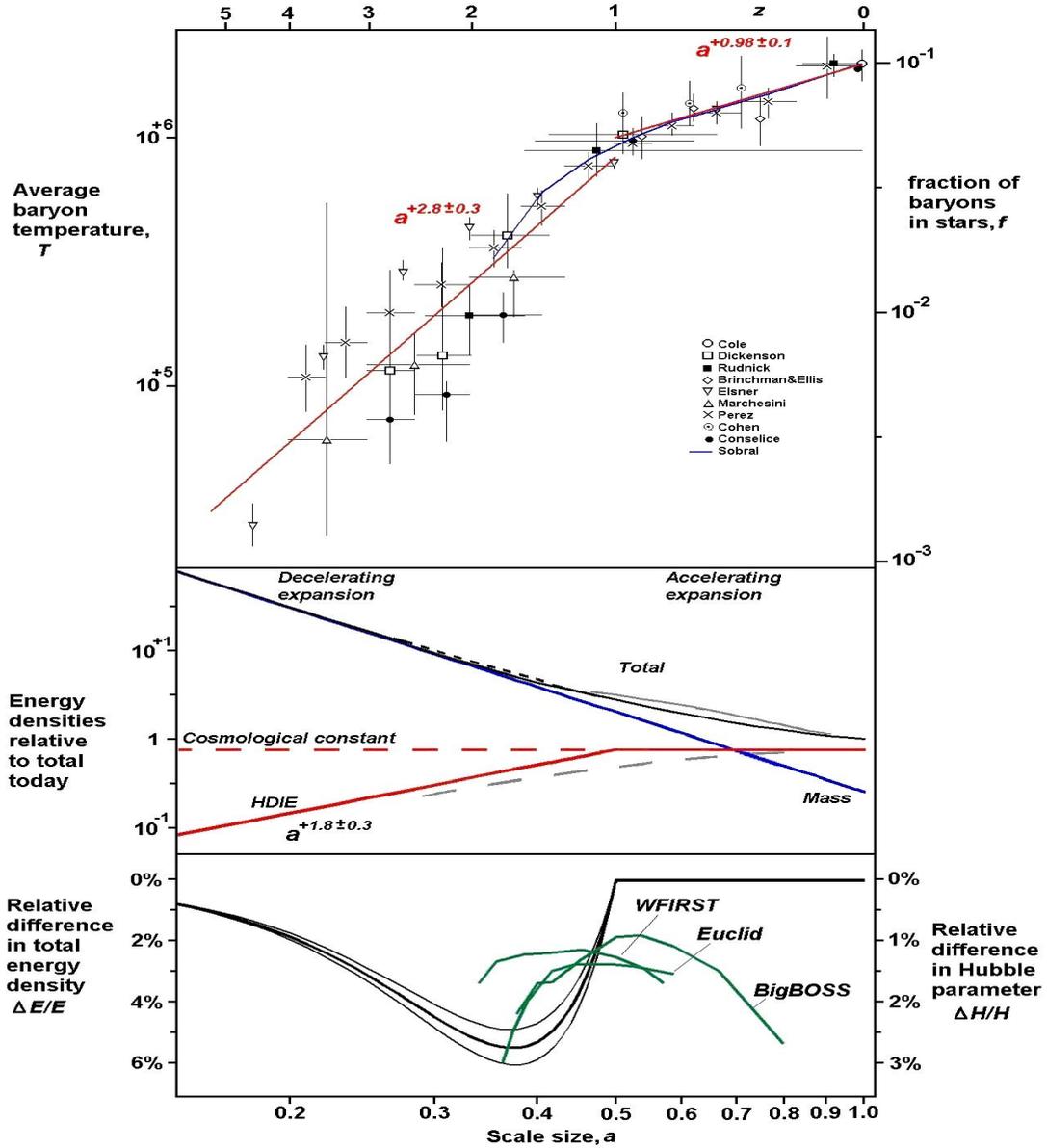

**Figure 1.** Plotted against log of universe scale size, *a*, and redshift, *z*, are three panels:
**(a) Upper Panel**: Log plot of measured average baryon temperature, *T*, and the fraction of all baryons in stars, *f* (various symbols and blue line: see text for measurement sources). Red lines- best power law fits to data points are $a^{+0.98\pm0.1}$ for $z < 1$, and $a^{+2.8\pm0.3}$ for $z > 1$.
**(b) Middle Panel:** Log plot of energy density contributions: red continuous line, HDIE energy density corresponding to the red line fit in the upper panel; dashed red line, cosmological constant; blue line, mass; solid black line, total for HDIE case; dashed black line, total for the case of a cosmological constant; grey dashed line, to illustrate problem of two parameter dynamic *w* model approximation (see text); grey continuous line, the gedanken experiment considered in section 3.2.
**(c) Lower Panel:** Linear plot of relative differences in total energy, and in Hubble parameter, between the HDIE model and a cosmological constant. The resolving thresholds of three next generation space and ground based measurements are shown for comparison in green.

Figure 1(a) shows a distinct change in power law around $z \sim 1$ and data either side of $z = 1$ are therefore considered separately. Applying linear least squares curve fitting to logarithmic values of the data in the redshift range $z < 1$ we observe a temperature gradient of $a^{+0.98\pm0.1}$. Then, assuming the baryon information bit number, $N$, scales as $a^{+2}$ from the holographic principle, total HDIE energy for $z < 1$ scales as $a^{+2.98\pm0.1}$. Since universe volume increases as $a^3$, there is a nearly constant HDIE energy density at $z < 1$. The HDIE equation of state then lies in the narrow range $-0.96 > w_{HDIE} > -1.03$ which includes the specific value, $w_{DE} = -1$ and thus satisfies dark energy requirement 2.

**2.2 Dependence on the Holographic principle**
HDIE can account for today's high dark energy value (requirement 1) solely by applying proven physics i.e. without requiring the holographic principle. Measurements of the present average baryon temperature (figure 1(a) right-hand axis intercept [34-43]), are combined with estimates of the information (entropy) associated with stellar gas and dust [30][31], and experimentally proven [25][26] Landauer's principle. However, for HDIE to account for the constant dark energy density $z < 1$, (requirement 2), the measured average baryon temperature relation ( T α $a^{+0.98\pm0.1}$ at $z < 1$, figure 1 (a)) has to be combined with the, as yet unproven, holographic principle relating information content to bounding area, $N$ α $a^{+2}$.

Now the Bekenstein-Hawking description [28] of a black hole, with entropy proportional to surface area, is widely accepted physics. But black holes exist at the maximum entropy holographic bound while the universe is some 30 orders of magnitude below the holographic bound. The holographic principle, whereby all 3-D space can be translated into a 2-D representation [15][16][44], is directly related to string theory and M-theory. Strong support for the holographic principle has been provided by a specific quantum theoretical example from string theory which allows for a holographic translation between one particular multidimensional space with gravity and another space with one less dimension but without gravity - the 'Maldacena duality' or 'anti-de-Sitter/conformal field theory' (AdS/CFT) correspondence [45]. Another theoretical work [46] effectively combines the holographic principle with Landauer's principle as in this present work, and suggests that gravity may emerge as an 'entropic force'. Note that our present work may be thought of as considering 'entropic energy'. Relevant theory in support of the holographic principle is still being developed and, as yet, there is no experimental proof of the principle. Attempts to directly verify the holographic principle by experiment are difficult and sometimes controversial [47]. Therefore the holographic principle, naturally extending the $N$ α $a^{+2}$ relation to all objects including those well below their maximum entropy, remains an attractive but unproven hypothesis, and thus accounts for the main speculative aspect of the HDIE model.

Nevertheless, the measurements plotted in figure 1(a), showing T α $a^{+0.98\pm0.1}$ at $z < 1$, closely centered round the T α $a^{+1}$ relation required for the HDIE explanation, provide significant support for the HDIE model. Accordingly, we continue our work below by considering HDIE primarily from a phenomenological point of view, and limit ourselves to only employing the main proposition of the holographic principle: i.e. that $N$ α $a^{+2}$.

**2.3 Dark energy predictions for $z >1$**
Figure 1(a) shows that the temperature gradient was much steeper at redshifts z > 1, with a wider spread in measured data points and a best power law fit of $a^{+2.8\pm0.3}$. Clearly HDIE energy density was increasing in this earlier period up to the point around $z \sim 1$ where HDIE leveled out at a near constant value that we showed above can account for both dark energy requirements from that time onwards.

In the following analysis we therefore assume that the level HDIE energy density $z < 1$ indeed accounts for all dark energy, and thus is located at a value three times the present mass energy density equivalent. Then figure 1(b) shows the mass energy density falling as $a^{-3}$ (blue line), the resulting HDIE energy density contribution with the above assumption (red line), and a cosmological constant for comparison (red dashed line). The $a^{+2.8\pm0.3}$ temperature dependance, $z > 1$, corresponds to an HDIE energy density gradient of $a^{+1.8\pm0.3}$ when information bit quantity, $N$, is again assumed to scale as $a^2$ from the Holographic principle. Then the mean $a^{+1.8}$ HDIE energy density variation corresponds to an equation of state $w_{HDIE} = -1.6$ for $z > 1$.

In figure 1(b) we also compare the total energy density from HDIE plus mass (black continuous line) with the total energy density of a cosmological constant plus mass (black dashed line). At first sight the two total energy curves lie very close with little apparent difference because they are plotted on a multi-decade log versus log plot.

Accordingly, in figure 1(c) the relative difference in total energy density, $\Delta E/E$, between HDIE plus mass and a cosmological constant plus mass is shown on a linear versus log plot. The lower, average and upper limits of the HDIE energy density gradient, $a^{+1.8\pm0.3}$, correspond to relative differences in total energy, $\Delta E/E$, in figure 1(c) that peak at −4.2%, −5.2%, and −6.2% respectively near $z \sim 1.6 - 1.7$. Although there is a clear change in gradient around $z \sim 1$ evident in the data points of figure 1(a), our fitting to gradients that change precisely at $z = 1$ may provide an overemphasized sharp transition in $\Delta E/E$ at $z = 1$. However, this transition should not significantly affect the size or the location of the predicted negative peak in $\Delta E/E$ at $z \sim 1.7$. At earlier times, $z > 4$, the higher mass density swamps any difference between HDIE and a cosmological constant. Later, as the massdensity falls $\Delta E/E$ begins to reflect the difference in the energy densities of the two dark energy components, peaking at $z \sim 1.7$ as HDIE energy density rapidly increases as $a^{+1.8\pm0.3}$ towards $z \sim 1$, after which time there is no difference between models.

While the Hubble constant, $H_0$, is the fundamental relation between the recessional velocities of objects in the universe and their distance from us today, $H_0$ is just the present value of the more general Hubble parameter, $H$. The Hubble parameter, $H$, varies with changes in universe expansion rate over time and is therefore a function of universe scale factor, $a$. Since total energy density, $E$, is proportional to $H^2$, (from the Friedmann equation, [48]) these three curves then correspond to relative differences in Hubble parameter, $\Delta H/H$, that peak at −2.1%, −2.6% and −3.1% respectively. The HDIE model thus predicts that the Hubble parameter around $z \sim 1.7$ should be $2.6 \pm 0.5\%$ less than that expected for a cosmological constant explanation for dark energy.

**2.4 Measurement capabilities of next generation instruments**
The HDIE predicted ~ 2.6% difference in the Hubble parameter can not be resolved by today's instruments which still have typical resolutions > 5% (See, for example, recent BOSS results, figure 21 of [49]). Fortunately the next generation of space and ground based dark energy instruments should be capable of making such a measurement. The future European Space Agency Euclid spacecraft [50], and the planned NASA WFIRST spacecraft [51] will both cover the redshift range $0.7 < z < 2.0$, while the ground based BigBOSS [52], LSST [53] and Dark Energy Survey [54] measurement campaigns will measure $z < 1.7$, $z < 5$ and $z < 2$, respectively. These experiments employ a combination of techniques: weak gravitational lensing to measure the growth of structure; supernova distances at low z; and baryon acoustic oscillations at higher z. The resolving limits of three of these dark energy experiments are shown in figure 1(c) (green lines) for comparison. The 2 − 3% difference in Hubble parameter around $z \sim 1.7$ should be resolved by all three experiments. At the time of writing these next generation measurement development timescales are: Dark Energy Survey starting a five year survey in 2012, BigBOSS first light 2016 with full science starting 2017; ESA Euclid launch 2019; LSST first light 2020 with full science starting 2022; and NASA WFIRST launch 2022.

Note that HDIE effectively provides a dynamic equation of state but, rather than a smooth variation, there are two distinct regimes: $w_{HDIE} = -1$ for $z < 1$; and $w_{HDIE} \sim -1.6$ for $z < 1$. Now it is usual when designing these experiments to characterise any dynamic equation of state, $w(a)$, by a smoothly varying two parameter model, typically given as : $w(a) = w_p + w_a(1 - a)$, where $w_p$ is the present value, the early value was $w_p + w_a$, and the mid-point transition occurs at $a = 0.5$, or $z=1$. The experimental figure of merit is then determined by how small the error ellipse is in the $w_p - w_a$ plane. For example, the ESA Euclid measurement accuracies equivalent to 1 sigma error are expected to be 0.02 in $w_p$, and 0.1 in $w_a$ up to $z \sim 2$ [50]. These accuracies are clearly sufficient to falsify HDIE where the nearest equivalent parameter values are $w_p=-1$ and $w_a=-0.6$, as compared to the cosmological constant values of $w_p = -1$ and $w_a = 0$. Note that this form of $w_p-w_a$ data analysis is not the ideal for HDIE because of the difference between such a smooth variation (grey dashed line in figure 1(b) fitted to the high gradient limit, $a^{+1.8+0.3}$, for emphasis) and the more distinct transition expected at $z \sim 1$ from HDIE. Rather than a single cluster of measurement data points on the $w_p-w_a$ plane HDIE predicts two clusters: one at $w_p=-1$ and $w_a=0$ independent of scale size, $a$, (i.e. $w_a = 0$) over that range $z > 1$. Thus the more appropriate mode of data analysis to identify any signature of HDIE is by a determination of the Hubble parameter, $H$, as a function of scale factor, $a$, or redshift, $z$ ( as in figure 21 of [49]).

The HDIE model is therefore falsifiable [55] since a failure to observe its predicted specific signature would clearly exclude this model. Note that there is an inherent lack of symmetry in falsification arguments. Although a positive observation of this signature would exclude a cosmological constant, it would not necessarily exclude all other models. For example, some form of quintessence field might produce a similar signature to that described below for HDIE, but then that model would also need to be equally capable of explaining the specific form of that observed signature.

## 2.5 Characteristic energy

The characteristic energy of HDIE, the energy equivalence of a bit of information, $k_B T \ln 2$, depends solely on temperature, $T$. Today, some 10% of the baryons are located in stars at temperatures $\sim 2 \times 10^7$K with characteristic energies $\sim 10^3$eV. As the remaining 90% of baryons exist at very much lower temperatures the average baryon temperature of baryons is , $\sim 2 \times 10^6$K, one tenth of the stellar temperature, corresponding to an average characteristic energy $\sim 10^2$eV.

We wish here to consider the characteristic bit energy of the 90% of baryons not involved in star formation and must first associate a representative temperature. We might consider the radiation temperature, $T'$, that would have the same energy density as matter: $\rho c^2 = \sigma T'^4$, where $\rho$ is the universe total mass density (including dark matter), and $\sigma$ the radiation constant. Substituting the radiation constant by its definition in terms of fundamental constants, we obtain the characteristic bit energy, $E_{char} = k_B T' \ln 2 = (15\rho\hbar^3 c^5/\pi^2)^{1/4} \ln 2$.

This definition was previously identified [56][57] as being identical to the characteristic energy of a cosmological constant (taking $\ln 2 \sim 1$, identical to equation 17:14 of [58]). In this way we obtain a value $T' \sim 35$K, corresponding to a characteristic energy of $\sim 3 \times 10^{-3}$ eV. Note that we expect a temperature around 10 times the temperature of the cosmic microwave background, CMB, since present matter energy density is $\sim 10^4$ times the CMB energy density.

Thus the, otherwise difficult to account for, low characteristic milli-eV energy usually associated with the cosmological constant [58] may be finally explained as an information bit equivalent energy. While the characteristic energy of HDIE corresponds to information concerning stars and star formation, the characteristic energy of the cosmological constant corresponds to information concerning those parts of the universe not involved in star formation. HDIE characteristic bit energy increases as $a^{+1}$ with increasing star formation while the cosmological constant characteristic energy falls as $\rho^{+1/4}$, or $a^{-3/4}$, with the cooling universe majority. Then it is difficult to see how the cosmological constant, with a characteristic energy falling as $a^{-3/4}$, can produce a total energy that increases as $a^{+3}$ as required for a constant energy density. In contrast, HDIE characteristic bit energy increases as $a^{+1}$ and total bit number, $N$, increases as $a^{+2}$ by the holographic principle to provide the required $a^{+3}$ total energy variation.

## 3 Gedanken experiment

We have used Landauer's principle above to show that the HDIE model may be considered a serious contender to explain dark energy. Landauer's principle can also explain Maxwell's Demon, the famous gedanken (thought) experiment of physics [23][27]. We continue here with what might be considered a less serious thought experiment, but one that nevertheless provides an interesting, information related, aspect to the universe's accelerating expansion.

### 3.1 Hypothetical computer simulation

Consider the amount of information that a hypothetical super computer located outside the universe would need to fully simulate the universe's baryons. We conveniently ignore how such a computer can be located outside the universe, how this information would be gathered, and any measurement limitations imposed by the uncertainty principle. For a full physics simulation we require that each spatial parameter of every baryon be registered to the maximum physically meaningful accuracy, i.e. at the resolution of the Planck length, $l_p = 1.6 \times 10^{-35}$m. Intergalactic baryons in the present universe, size $l_u \sim 10^{27}$m, will then require an accuracy of one part in $6 \times 10^{61}$ ($\sim 2^{205}$) and hence require 205 bits per spatial parameter. Similarly baryons located in giant molecular clouds, size $l_{gmc} \sim 10^{18}$m, and baryons located in typical stars, e.g. the sun size $l_s \sim 10^9$m, require ~175 bits and ~145 bits respectively per spatial parameter.

Rather than consider the total information required for our simulation it is more convenient to limit ourselves to estimating the average number of bits per spatial parameter per baryon. We assume that the 10% of baryons presently located in stars that now require ~145 bits per parameter were, at earlier times, located in giant molecular clouds that required ~30 bits more at ~175 bits per parameter. We model this change from giant molecular clouds to stars from the variation of the fraction, $f(a)$, of baryons in stars as a function of scale size using the power law fits to the measurements plotted in figure 1(a). Meanwhile the remaining 90% of baryons, intergalactic baryons not involved in star formation, increased as $\log_2(a)$ up to their present 205 bits per parameter. The average number of bits per parameter per baryon, $n_{av}$, is a function of scale size, $a$, given by: $n_{av} = (1 - f(1)) \log_2(a\, l_u/l_p) + (f(1) - f(a)) \log_2(l_{gmc}/l_p) + f(a) \log_2(l_s/l_p)$.

Inserting values of $f(a)$ from figure 1(a) we find that $n_{av}$ increased with intergalactic baryons but reached a peak value of 200.03 bits at $a \sim 0.32$, but then decreased due to increasing star formation to today's value of 199.02 bits, almost exactly one bit below the peak value. This one bit loss can be explained by the 10% of baryons that formed stars lost 30 bits per spatial parameter, contributing a loss of 3 bits to $n_{av}$, while the 90% intergalactic baryons only added 2 bits to $n_{av}$ between $a = 1/4$ and the present, $a = 1$.

The amount of information required to simulate an independant system should never decrease, otherwise it must imply a decrease in that system's number of states or information, contrary to the 2nd law. Now, if the universe expanded faster to double its expected size over the recent period, this would increase the contribution of the 90% intergalactic baryons to $n_{av}$ by a further 1 bit. Then we could effectively compensate for our loss of one bit in $n_{av}$ and satisfy the 2nd law again. Interestingly, dark energy has indeed doubled the size of the universe, exactly as we require, since it has increased the energy density by a factor of four, corresponding to a doubling of the Hubble parameter. Figure 1(b) grey continuous line, uses the above relation and assumptions to show the minimum required variation in total energy density that ensures there is no decrease in the amount of information required as input to our simple computer simulation during this period. This variation can be seen to lie close to that deduced from the effects of dark energy (whether due to HDIE or a cosmological constant).

It is a surprise to find that the accelerating expansion was necessary for the universe to comply with the 2nd law and ensure that there was no decrease in the amount of information required as input to our simple thought experiment! Note that while the approach here is based on just a few simple assumptions they are all none the less quite reasonable. For example, many of the 90%, intergalactic baryons, not involved in star formation, do not move freely throughout the whole universe but are probably constrained to intergalactic filaments whose dimensions presumably stretch with the increasing space between galaxies and hence still have dimensions that scale with $a$. We have ignored the information represented by CMB as it has remained near constant since decoupling with CMB wavelength increasing in proportion to universe size. We have also ignored those baryons still in giant molecular clouds, yet to take part in star formation, but these will just add a constant amount to $n_{av}$. If we had used a different minimum resolution, for example the Fermi length, $10^{-15}$ m, the above bit numbers would be 66 bits less but, without the doubling of universe size from dark energy, there would still have been the same reduction of 1 bit in $n_{av}$. So, although there is considerable uncertainty in absolute quantity of information required for our simulation, we can reasonably say that the doubling in universe size due to dark energy was just what was required to ensure that the amount of information needed as input to our computer simulation did not decrease.

Of course it is still possible that our requirement for about one bit is just because we chose giant molecular clouds (size $\sim 10^{18}$m) as the starting points for star formation. For comparison, at the two extremes of starting point either side, we would have obtained a value close to two bits drop in $n_{av}$ if we had considered star formation as starting all the way from the parent galaxies (size $\sim 10^{21}$m) or a value close to zero change in $n_{av}$ if we had considered that star formation only started much later, at the final pre-stellar stage of proto-stellar nebula (size $\sim 10^{15}$m). However, given the typical star formation sequence and the timescale considered in figure 1, it seems most reasonable to consider pre-existing giant molecular clouds as the effective beginning points for star formation.

### 3.2 Algorithmic information content
The relation between this simulation information and the actual information intrinsic to the universe is analogous to the relation between the algorithmic information content (algorithmic entropy or Kolmogorov complexity), the size of the smallest algorithm that can generate a dataset and the actual amount of information contained within that dataset [59,60]. For non-random datasets algorithmic information is always less than the information in the dataset. For a truly random dataset, i.e. one that can not be calculated by an algorithm, the algorithmic information is always just slightly greater than the amount of information in the dataset. At a minimum it is greater by the size of the small program required to access the random dataset that must then be completely included as data, constituting the bulk of program code.

At $\sim 200$ bits per spatial parameter each of the $\sim 10^{80}$ baryons in the universe requires $\sim 10^3$ bits, giving a total baryon simulation requirement $\sim 10^{83}$ bits. Note that this value is not very dependent on simulation resolution, whether say at Planck or Fermi lengths. Then, by analogy to algorithmic information content, we see that this simulation requirement is, as expected, less than the above $N \sim 10^{86}$ bits of HDIE because significant structure, or non-randomness, exists in the form of galaxies, stars etc. We can not deduce much from the actual size of this difference because of both the uncertainties in entropy estimation and the simplicity of our thought experiment argument. However, in future, the maximum rate of increase of simulation information is limited to the slow $\log_2 a$ rate of simulated intergalactic baryons while holographic information should continue to increase at the much faster rate of $a^2$. Then we should expect a growing significant difference that must further reflect the evolving level of non-randomness in the universe caused by increased structure formation.

### 4 Implications for the cosmos
The information based approach followed in this work leads directly to several implications for the cosmos, especially if the predicted HDIE model signature is observed, and HDIE thus found to be the correct explanation for dark energy.

The first implication concerns the reason why the temperature variation $a^{+0.98\pm 0.1}$ so closely follows $a^{+1}$ since $z\sim 1$ to provide the near constant HDIE energy density, $-0.96 < w_{HDIE} < -1.03$. If star formation had continued to proceed at the earlier faster rate, then it would have continued the steep $a^{+2.8\pm0.3}$ average baryon temperature increase after $z\sim1$. This would have increased HDIE dark energy well above its present value, lead to much greater acceleration and greater expansion, but in turn, would have resulted in much less star formation. It would appear that since $z\sim1$ there has been a balance, or feedback, between expansion acceleration and star formation that has naturally maintained the star formation rate close to $a^{+1}$ for a constant dark energy density. Note that the reduced rate of star and structure formation starting at $z\sim1$ was previously attributed to the onset of acceleration [61]. Thus HDIE provides a natural explanation for the reason why $w_{DE}=-1$ since $z\sim1$.

The second implication concerns the cosmic coincidence problem. Our existence just now in the era dominated by dark energy is considered an unlikely coincidence. However, HDIE dark energy density increased with increasing entropy and increasing baryon temperature while mass density decreased with increasing universe scale size. There had to be a time when HDIE energy density reached a level comparable to mass energy density to initiate acceleration (provided that time was reached before $f(a)=1$). Similarly, the likelihood of our existence also increased as overall star formation increased, and thus more likely to occur after HDIE started to make a significant contribution to the universe energy budget, effectively removing the cosmic coincidence problem.

The third implication concerns how long the present period of accelerating expansion will last. Acceleration will continue provided that the overall universe equation of state, $w<-1/3$ [9]. This threshold corresponds to HDIE energy density falling off as $a^{-2}$, and, assuming the total information, $N$, continues to follow the Holographic principle as $a^{+2}$, provides a limiting average baryon temperature, $T$, variation of $a^{-1}$. Thus, acceleration due to HDIE will continue providing $T$ does not fall off more steeply than $a^{-1}$. Computer simulations of future average baryon temperatures, $T$, up to $a=200$ [62], predict a leveling off of $T$ since $f(a)$ is limited by definition to $f(a) < 1$, with a slow eventual fall as star formation ceases, but falling less steeply than the threshold gradient of $a^{-1}$. Thus acceleration should continue, until at least the universe has increased in size by a factor of 200.

Clearly the fourth implication is that, should the predicted signature of HDIE be observed, it would provide very strong support for the holographic principle (see section 2.2).

The final implication concerns how the universe as a whole still manages to satisfy the 2nd law when degrees of freedom are lost as matter becomes denser when stars are formed. It has been suggested [63] that the loss of thermodynamic entropy due to structure and star formation is counteracted by a gain in gravitational entropy. However, our simple gedanken experiment above implies that the extra expansion from dark energy acceleration provides enough of an increase in inter-galactic states to compensate for those states lost during star formation. Interestingly, with the HDIE explanation for dark energy, the extra expansion is itself a direct result of star formation.

## 5 Summary

Computer scientist Landauer [64] emphasized that "Information is Physical" and astrophysicist Wheeler [65] went further, declaring with his famous slogan "It from Bit", that information may be more fundamental than matter. All of the arguments put forward in this paper for the HDIE dark energy explanation, as well as those used in the above thought experiment, also combine to point to the importance of considering information as one of the fundamental properties of the universe.

Most importantly, we have shown that HDIE can account for dark energy both qualitatively and quantitatively, accounting for both key dark energy properties in the redshift range $z < 1$: the constant dark energy density and that energy density value. Furthermore, with the HDIE explanation for dark energy we no longer have the coincidence problem.

At higher redshifts, HDIE should produce a clear signature, predicting that at $z$~1.7 the Hubble parameter will have a value 2.6±0.5% less than that expected for a cosmological constant. Then the HDIE model is falsifiable as the size and location of this predicted signature lies within the resolvable ranges of the next generation of dark energy measurements.

**References**
[1] A.G. Riess et al., Astronomical J. 116 (1998) 1009.
[2] S. Perlmutter et al.,Astrophys. J. 517 (1999) 565.
[3] A.G. Riess et al.,Astrophys. J. 659 (2007) 98.
[4] D.J. Bacon, A.R. Refregier and R.S. Ellis,Mon. Not. R. Astron. Soc.318 (2000) 625.
[5] F. Beutler, et al.,Mon. Not. R. Astron. Soc.416 (2011) 3017.
[6] B.D Sherwin, et al. Phys. Rev. Lett.107 (2011) 021302.
[7] C. Blake, et.al. Mon. Not. R. Astron. Soc.415(2011) 2876.
[8] S.M.Carroll, Carnegie Observatories Astrophysics Series, Vol 2, Measuring and Modelling the universe, ed. W.L.Freedmann Cambridge University Press (2003).
[9] J.A. Frieman, M.S. Turner, D. Huterer,Ann. Rev. Astron. Astrophys.46 (2008) 385.
[10] M. Tegmark, et al.,Phys. Rev. D 74 (2006) 123507.
[11] S. Weinberg,Rev. Mod. Phys, 61 (1989) 1.
[12] M. Li, Phys. Lett. B 603 (2004) 1.
[13] Y. Gong, B. Wang and Y-Z Zhang,Phys. Rev. D 72 (2005) 043510.
[14] A. Shalyt-Margolin, Entropy 12 (2010) 932.
[15] W. Fischler and L. Susskind,arxiv: hep-th/9806039 (1998).
[16] G. t' Hooft, Stud. Hist. Phil. Mod. Phys. 32 (2001) 157.
[17] Y.Gong and Y-Z Zhang Y-Z, Class. Quantum Grav. 22 (2005) 4895.
[18] J-W Lee, J Lee and H-C Kim,JCAP08 (2007) 005.
[19] R. Landauer, IBM J. Res. Dev. 3 (1961) 183.
[20] M.P. Gough, Entropy 13 (2011) 924.
[21] B. Piechocinska, Phys. Rev. A 61 (2000) 062314.
[22] R. Landauer, Nature 335 (1988) 779.
[23] H.S. Leff, A.F. Rex, Eds. Maxwell's Demon 2: Entropy, Classical and Quantum Information, Computing IOP Publishing Ltd: London, UK, (2003).
[24] M. Hilbert, P. López,Science 332 (2011) 60.
[25] S. Toyabe, et al., Nature Physics 6 (2010) 988.
[26] A. Berut, et al., Nature 483 (2012) 187.
[27] K. Maruyama, F. Nori, V. Vlatko, Rev. Mod. Phys. 81 (2009) 1.
[28] J.D. Bekenstein, Phys. Rev. D 7 (1973) 2333.
[29] L. Susskind, J. Math. Phys. 36 (1995) 6377.
[30] P.H. Frampton, S.D.H. Hsu, S.D.H, D.Reeb, T.W. Kephart, Classical Quantum Gravity 26 (2009) 145005.
[31] C.A. Egan, C.H. Lineweaver, Astrophys. J. 710 (2010) 1825.
[32] R.Buosso, R. Harnik, G.D.Kribs, G.Perez, ArXiv hep-th/0702115, (2007).
[33] R. Ruffini, J.A. Wheeler, Physics Today 24 (1971) 30.
[34] S. Cole, et al., Mon. Not. R. Astron. Soc. 326 (2001) 255.
[35] M. Dickinson, C. Papovich, H.C. Ferguson, T. Budavari, Astrophysical J. 587 (2003) 25.
[36] G. Rudnick, et al, Astrophys. J. 599 (2003) 847.
[37] J. Brinchmann, R.S. Ellis, Astrophys. J. 536 (2000) L77.
[38] F. Elsner, G. Feulner, G., U. Hopp, Astronomy and Astrophysics 477 (2008) 503.
[39] D. Marchesini, et al, Astrophys. J. 701 (2009) 1765.
[40] P.G. Perez-Gonzalez, et al., Astrophys. J. 675 (2008) 234.
[41] J.G. Cohen, Astrophys. J. 567 (2002) 672.
[42] C.J. Conselice, et al, Astrophys. J. 620 (2005) 564.

[43] D. Sobral, et al, arxiv: 1202.3436 (2012).
[44] R. Buosso, Rev. Mod. Phys. 74 (2002) 825.
[45] J. Maldacena, arxiv:hep-th/9711200v3, (1998).
[46] E. Verlinde, arxiv:hep-th/1001.0785v1, (2010).
[47] A. Cho, Science 336 (2012) 147.
[48] A. Friedman, General Relativity and Gravitation 31 (1999) 1991.
[49] N.G. Busca, et al, arxiv: 1211.2616v1 (2012).
[50] Euclid Definition Study Report ESA SRE 12 July 2011, sci.esa.int/science-e/www/object/index.cfm?fobject=48983 (2011).
[51] NASA WFIRST Final Report jdem.gsfc.nasa.gov/science/sdt_public/WFIRST_SDT_Final_Report.pdf (2011).
[52] D.J. Schlegel, et al, arxiv: 0904.0468v3 (2009).
[53] Z. Ivezic, et al, LSST arxiv: 0805.2366v2 (2011).
[54] Dark Energy Survey proposal www.darkenergysurvey.org/reports/proposal-standalone.pdf (2012).
[55] K. Popper, The logic of scientific discovery Routledge, London, UK, (1959).
[56] M.P. Gough, T. Carozzi, A.M. Buckley, arxiv: astro-ph/0603084 (2006).
[57] M.P. Gough, Entropy 10 (2008) 150.
[58] P.J.E. Peebles, Principles of Physical Cosmology Princeton University Press, (1993).
[59] W.H. Zurek, Nature 341 (1989) 119.
[60] S. Devine, Entropy 11 (2009) 85.
[61] L. Guzzo, et al., Nature 451 (2008) 541.
[62] K. Nagamine, A. Loeb, New Astron. 9 (2004) 573.
[63] R. Penrose, The Road to Reality Jonathan Cape, London, UK, (2004).
[64] R. Landauer, Physics Today 44(5) (1991) 23.
[65] J.A. Wheeler, Complexity, Entropy, and the Physics of Information Addison-Wesley, Redwood City, California, US, (1990).